\documentclass[twocolumn,10pt]{asme2e}

\usepackage{graphicx}
\usepackage{caption}
\usepackage{subcaption}
\usepackage{lipsum}
\usepackage{amsmath}
\usepackage{multicol}

\confshortname{FEDSM 2020}
\conffullname{the ASME 2020 Fluids Engineering Division Summer Meeting}

\confdate{July 12-16}
\confyear{2020}
\confcity{Orlando, Florida}
\confcountry{USA}

\papernum{FEDSM2020-12388}

\title{The Effect of Varying Viscosity in Turbulent Channel Flow}

\author{Victor Coppo Leite
    \affiliation{
	Ken and Mary Alice Lindquist\\Department of Nuclear Engineering\\
	Pennsylvania State University\\
	University Park, PA 16802\\
    Email: vbc5085@psu.edu
    }
}

\author{Elia Merzari
    \affiliation{
	Ken and Mary Alice Lindquist\\Department of Nuclear Engineering\\
	Pennsylvania State University\\
	University Park, PA 16802\\
    Email: ebm5153@psu.edu
    }
}

\begin{document}

\maketitle

\begin{abstract}
{In this article we examine channel flow subject to spatially varying viscosity in the streamwise direction. The Reynolds number is imposed locally with three different ramps. The setup is reminiscent of transient channel flow, but with a space-dependent viscosity rather than a time dependent viscosity. It is also relevant to various applications in nuclear engineering and in particular in test reactors, where the viscosity changes significantly in the streamwise direction, and there is a severe lack of Direct Numerical Simulation (DNS) data to benchmark turbulence models in these conditions.

As part of this work we set up a novel benchmark case: the channel is extended in the stream-wise direction up to 20\(\pi\). The viscosity is kept constant in the first 4\(\pi\) region. This inlet region is used as a cyclic region to obtain a fully developed flow profile at the beginning of the ramping region. In the ramping region the Reynolds number is linearly increased along the channel. The flow is homogenous in the spanwise direction, while it is non-homogenous in the stream-wise and wall-normal direction. We perform here Direct Numerical Simulation (DNS) with Nek5000, a spectral-element computational fluid dynamics (CFD) code developed at Argonne National Laboratory.

In this study, specific focus is given to the investigation of turbulence properties and structures in the near-wall region along the flow direction. Turbulent statistics are collected and investigated. Similarly to transient channel flow, the results show that a variation in the Reynolds across a channel does not cause an immediate change in the size of turbulent structures in the ramp region and a delay is in fact observed in both wall shear and friction Reynolds number. The results from the present study are compared with a correlation available in the literature for the friction velocity and as a function of the Reynolds number.}
\end{abstract}

\begin{nomenclature}
\entry{$a$}{Viscosity linear coefficient.}
\entry{$C$}{Contraction parameter.}
\entry{${F_{Ci}}$}{Delayed function of Cases ${i}$=I,II and III.}
\entry{$R$}{Relaxation parameter.}
\entry{$Re$}{Reynolds number.}
\entry{$Re_{\tau}$}{Friction Reynolds number.}
\entry{$u_{\tau}$}{Friction velocity.}
\entry{$x$}{Streamwise direction.}
\entry{$y$}{Wall-normal direction.}
\entry{$z$}{Spanwise direction.}
\entry{$\delta$}{Height of the turbulence channel.}
\entry{${\delta}_{i}$}{Streamwise position where Region $i$ starts, ${i}$=I,II and III.}
\entry{${\Lambda}$}{Signal contribution in the delayed function.}
\entry{$\nu$}{Kinematic Viscosity.}
\entry{${\tau}_w$}{Wall Shear stress }
\end{nomenclature}

\section*{INTRODUCTION}
In various applications in nuclear engineering and in particular in test reactors, heat removal is carried by single-phase axial flow. In these applications, we observe sharp changes in molecular viscosity while the density presents very limited changes.  For example in a sodium reactor the viscosity drops by 30\% across the core \cite{finkthermodynamic} while the density drops by less than 5\%.  For test reactors operating with water at atmospheric pressure the drop in viscosity across the core can exceed 100\% while the density changes by only 2 \% (e.g., the values are computed assuming an average temperature increase across the core of 40 K).As a consequence, the Reynolds number increases often two-folds across the channel, with an inlet value often transitional.  In these conditions, turbulence changes significantly across the length of the channel with redistribution and thinning of the boundary layers.

While there have been numerous studies that investigate the interaction of temperature boundary layers with velocity boundary layers and changes in properties, the focus is often on the effect of buoyancy, which is not a primary concern for the cases under investigation. There is a relative derth of data, in particular using DNS, on the effect of changing viscosity over a spatially developing boundary layer. The present study is a step toward filling this  gap. The data generated will be used to benchmark Reynolds Averaged Navier Stokes turbulence models used for the design and safety evaluation of advanced reactors and test reactors.

While the ultimate objective is to simulate the effect on channel flow of a spatially developing temperature boundary layer, we start with an imposed  space-dependent viscosity (and consequently Reynolds number). This has the benefit of reducing complexity and separating the effect of changing viscosity from the problem of a spatially developing temperature boundary layer. We note the similarity of this approach to transient channel flow, in which the Reynolds number of a double-periodic channel flow simulation is changed in time.  Maruya et al.~\cite{maruyama1976} carried out one of the earliest studies on the effect of a step increasing of flow rate in a turbulent flow. Later, He et al.~\cite{he2000} and  Greenblatt \& Moss~\cite{greenblatt2004} imposed linearly increasing or decreasing excursions of flow rate in a turbulent flow. More recently, He \& Seddighi ~\cite{he2015} performed a much more extensive study on this topic using DNS through a series of transient flows with systematically varied initial and final Reynolds numbers. Interestingly, He \& Seddighi ~\cite{he2015}  described transient channel flow as a bypass transition. In the course of the manuscript we will see similarities between the case of interest and transient channel flow. We note however that a direct mapping through the Taylor hypothesis is not possible, in fact a spatial change of viscosity cannot be directly related to a time-dependent channel flow in which the conditions of the entire channel are changed at once.

As part of this work we set up a novel benchmark case: the channel is extended in the streamwise direction up to 20\(\pi\). The viscosity is kept constant in the first 4\(\pi\) region. This inlet region is used as a cyclic region to obtain a fully developed flow profile at the beginning of the ramping region. In the ramping region the Reynolds number is linearly increased along the channel. The flow is homogenous in the spanwise direction, while the flow is non-homogenous in the streamwise and wall-normal direction respectively. 

Three cases varying the linear rate of the Reynolds number are tested, for each one of these cases Direct Numerical Simulation is performed with Nek5000, a spectral-element computational fluid dynamics (CFD) code developed at Argonne National Laboratory. The simulations will be used to develop a full DNS dataset. Additionally, a Reynolds Averaged Navier Stokes (RANS) model was developed using the \(k-\tau\) turbulence model~\cite{speziale1992}. In the methods section we present the test cases, introduce Nek5000 and the convolution analysis. In the results section we compare simulation results with available data and we present RANS results considering varying viscosity. We focus in particular on spatially developing $Re_{\tau}$ as a function of $x$ and the difference observed with predictions based on correlations obtained for fully developed channel flow. Finally, we examine coherent structures and in particular how streaks develop in the streamwise direction. We observe strong similarities with what observed for transient channel flow for the spanwise correlation of the streaks. The results call into question the application of standard turbulence models and wall functions to cases with strongly varying viscosity, which should be evaluated for benchmarks like the one provided within this study.


\section*{METHODS - TEST CASES}

\begin{figure}[t]
	\centering
	\scalebox{0.5}
	{\includegraphics{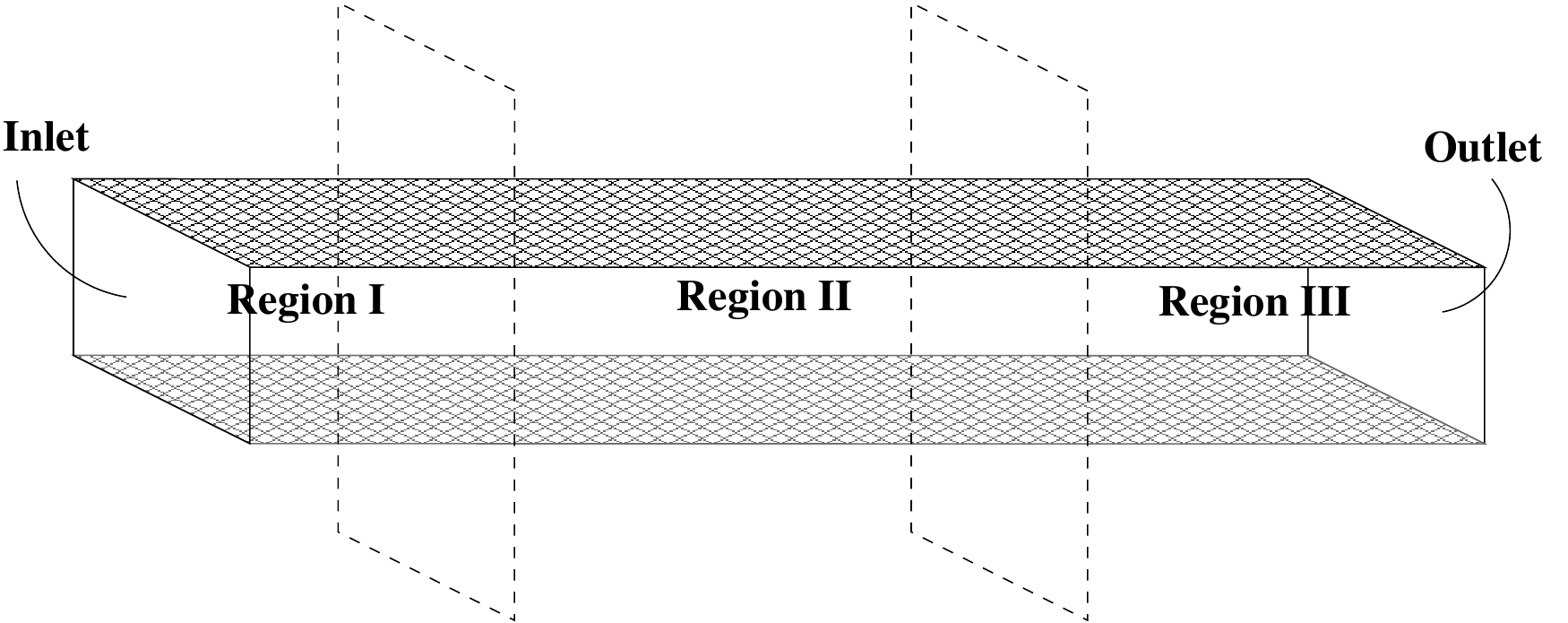}}
	\scalebox{0.4}
	{\includegraphics{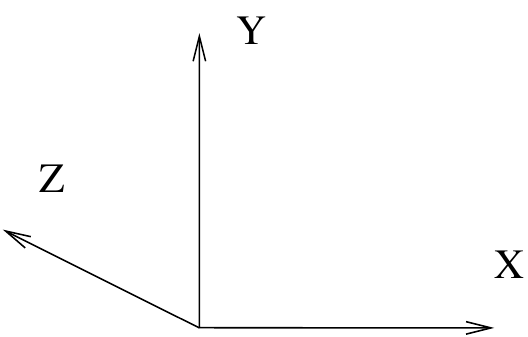}}
	\caption{Geometry of the turbulence channel, the channel is divided into three different Regions.}
	\label{fig:geometry_turbchannel}
	\end{figure}

\begin{figure}[t]
	\centering
	\scalebox{0.5}
	{\includegraphics{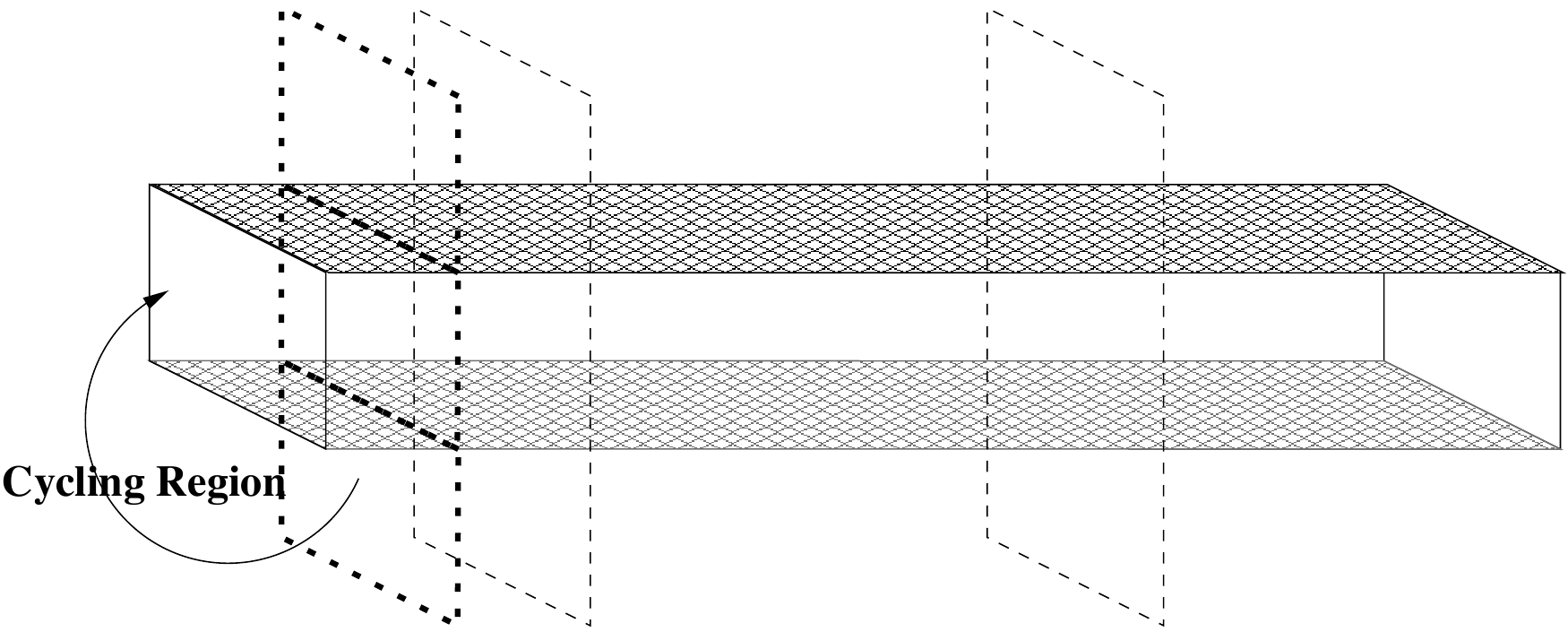}}
	\caption{Cyclic region in the inlet.}
	\label{fig:cycling_region}
\end{figure}

A schematic of the turbulence channels simulated are shown in Fig.~\ref{fig:geometry_turbchannel}. In this figure, Regions I, II and III are defined by two planes crossing the channel. A cycling region is implemented within Region I, Fig.~\ref{fig:cycling_region}. In this problem, the viscosity \(\nu\) is a function of \(x\), i.e., the streamwise direction. This parameter has constants values along Regions I and III respectively, while in region II it decreases with respect to the inverse of \(x\). In the present work, three cases varying the length of Region II are studied. Cases I, II and III have their Region II length of respectively \(16\pi\), \(8\pi\) and \(4\pi\).

\begin{figure}[t]
	\centering
	\scalebox{0.5}
	{\includegraphics{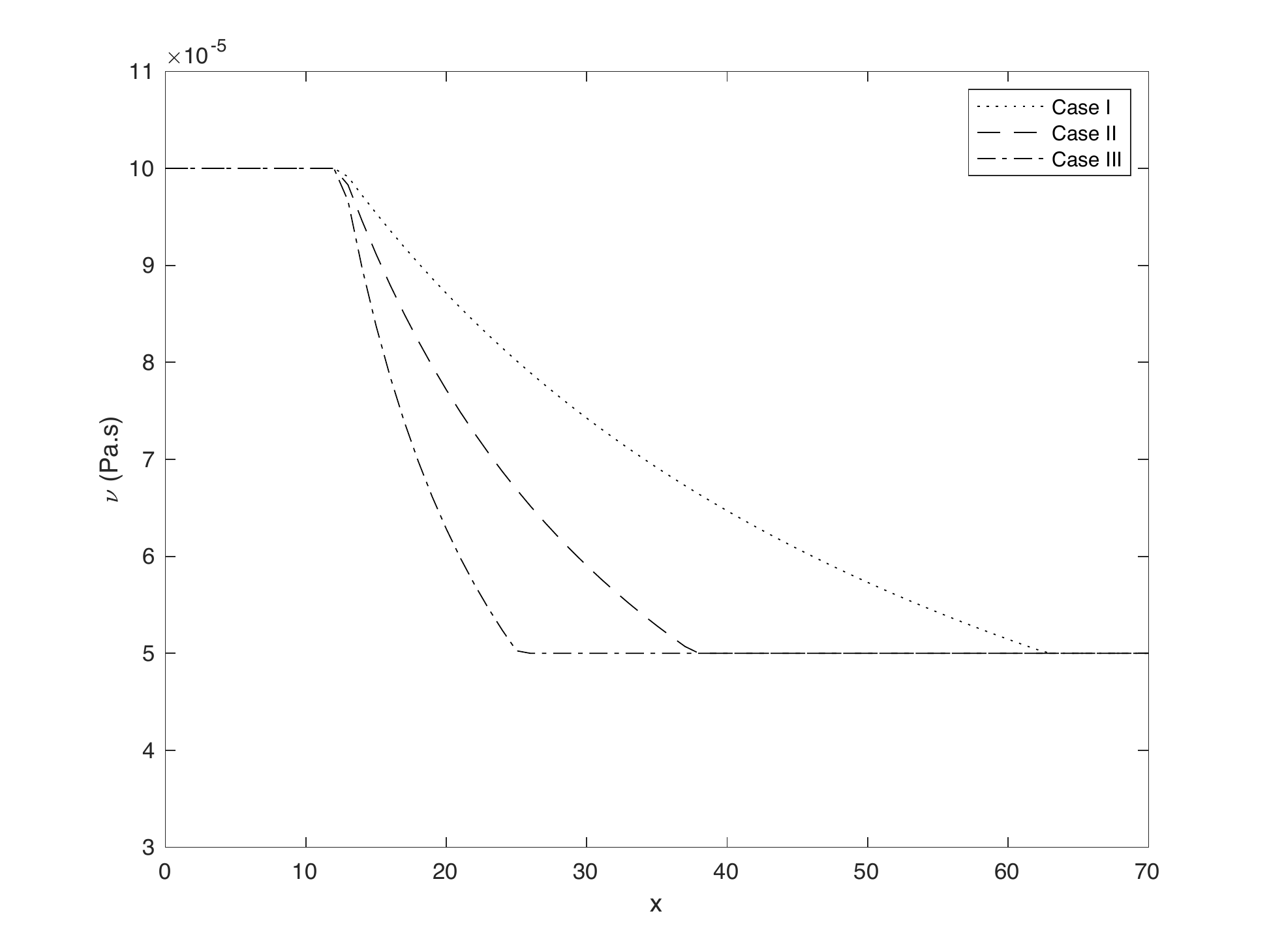}}
	\caption{The viscosity of the considered Cases as a function of the streamwise direction.}
	\label{fig:viscosity}
	\end{figure}

\begin{figure}[t]
	\centering
	\scalebox{0.5}
	{\includegraphics{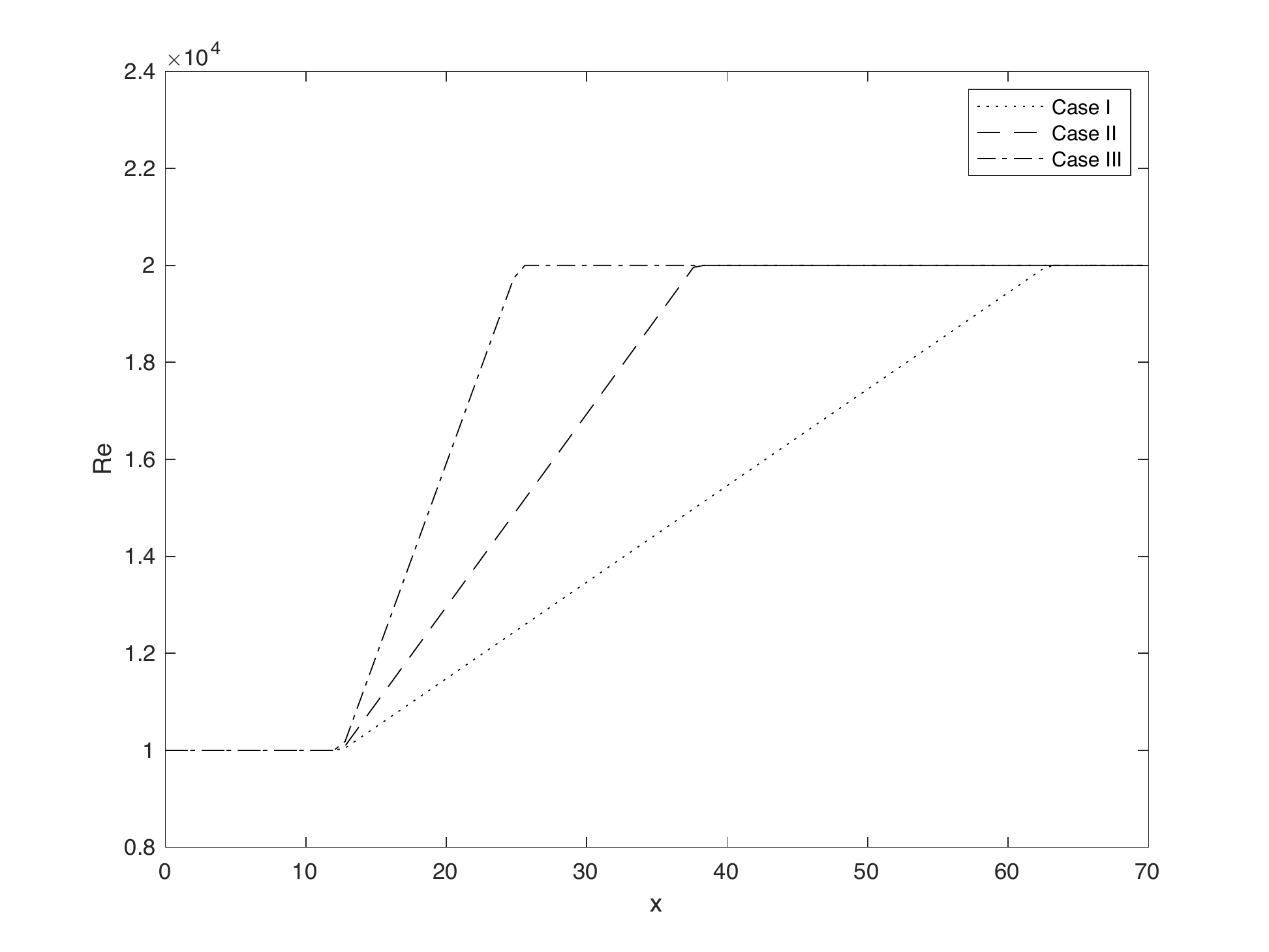}}
	\caption{The Reynolds number as a function of the \(x\) for the considered Cases.}
	\label{fig:reynolds}
\end{figure}

Fig.~\ref{fig:viscosity} shows the plot of the viscosity as a function of \(x\) for each one of these cases and Fig.~\ref{fig:reynolds} shows the plot of the Reynolds number for them. In all considered cases the range of the Reynolds number is \(10,000\) - \(20,000\) and it linearly increases in the streamwise direction, but with different ramps. For all three cases the inleat flow in Region I is considered to be fully developed with \(Re_{\tau}=550\), i.e., the same conditions as in \cite{hoyas2008}. Periodic condition is considered for the boundaries of the spanwise direction, i.e., \(z\) axis, and finally wall conditions are considered for the boudaries of the vertical direction, i.e., \(y\) axis.

\section*{METHODS - DIRECT NUMERICAL SIMULATION}

In order to resolve the finest turbulent scales, the calculations of this work has been developed through DNS.
These kind of simulations are able to resolve the finest turbulent length scales without using any turbulent model. Since the present work is focused on studying the contribution of the smaller scales to the energy cascade, it is required to use DNS rather than Reynolds Averaged Navier-Stokes (RANS) or Large Eddy Simulations (LES), although there is a substantial growth of the computational cost.

Nek5000 is employed, a spectral element code developed in Argonne National Laboratory (ANL).  In Nek5000 \cite{fischer2015nek5000} \cite{fischer1997} the domain is discretized in curvilinear hexahedral  or quadrilateral elements, conforming to the domain boundaries. Functions within each element are expanded using Lagrangian polynomials built on Gauss Lobatto Legendre collocation points and operators are cast in tensor-product form. The pressure can be solved at the same polynomial order of the velocity $N$ ($P_{N}-P_{N}$ formulation) or at lower order $N-2$ ($P_{N}-P_{N-2}$ formulation). Two time-stepping schemes, both up to third order, are available: BDF and OIFS \cite{fischer2003implementation}. The latter has the advantage of less severe stability limitation allowing for $CFL>1$, but is characterized by a larger cost per time step. We note that the temporal discretization is typically based on a high-order splitting that is third-order accurate in time.
Nek5000 features a stress-formulation solver capable of solving the three velocity components at once: this is essential for cases with non-aligned symmetry boundary conditions, variable viscosity cases and RANS.

The pressure substep requires a Poisson solver at each step, which is performed through multigrid-preconditioned GMRES iteration coupled with temporal projection to find an optimal initial guess. Particularly important components of Nek5000 are its scalable coarse-grid solvers that are central to parallel performance. For both weak scaling and strong scaling using Algebraic Multi Grid (AMG) for the coarse-grid solve is essential above 250,000 elements. Nek5000 employs a pure MPI parallel implementation. Nek5000 has received extensive validation in numerous references, we cite here \cite{merzari2013} and \cite{Obabko2011}.

\begin{figure}[t]
	\centering
	\scalebox{0.16}
	{\includegraphics{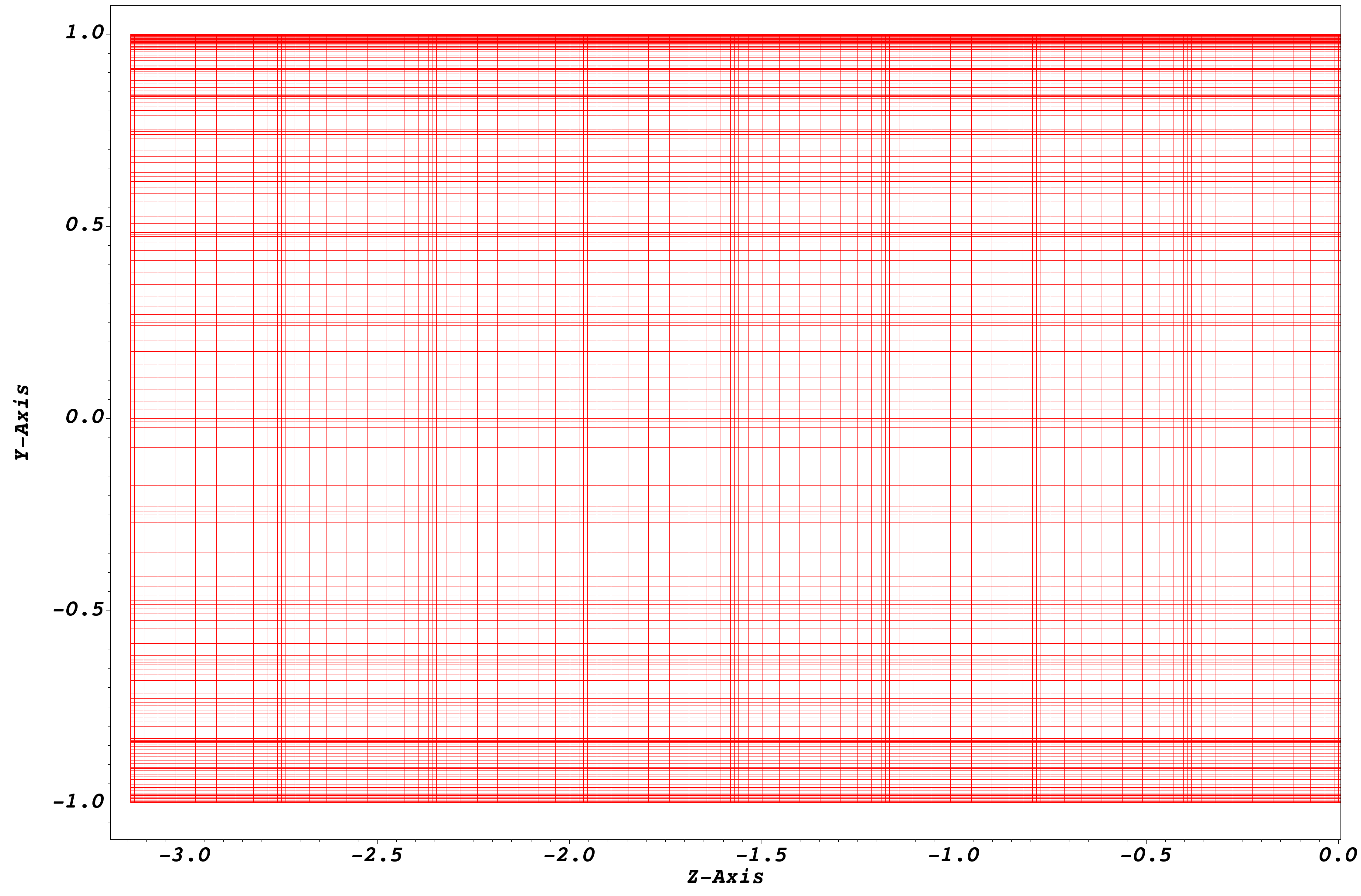}}
	\caption{The grid employed in the simulation from half of the channel's cross section. 11th order.}
	\label{fig:mesh}
\end{figure}

Lagrangian polynomials of up to the 15th degree have been employed to discretize the velocity field in this case. The total mesh count exceeds 100,000 hexahedral elements distributed homogeneously in the streamwise direction. Figure~\ref{fig:mesh} shows an example of the grid from half of the channel's cross section. One should notice that the discretization presented by this particular area is identical through all model's domain and it is only presented half of the cross-section for better visualization of the frame. The discretization has been developed  to match accepted standards for the DNS of channel flow. This includes: $y^{+}<1$ near wall, 10 points below $y^{+}<10$, $\Delta x^{+}<4$ and $\Delta z^{+}<8$.

\section*{METHODS - CONVOLUTION ANALYSIS OF THE FRICTION REYNOLDS NUMBER}

The friction Reynolds number has been calculated via numerical simulations for Cases I, II and III and this result was compared to the values obtained using an existing expression from Ref.~\cite{pope} valid for fully developed turbulent flows. A delay in space can be seen between \(Re_{\tau}\) when comparing the simulations' results with the values yielded from the mentioned expression. This way, these results are treated as signals in the space and a convolution operation is performed over the analytical expression for fully developed turbulent flows in order to built a function that should matches with the obtained values from the simulations.

To calculate the friction Reynolds number from the simulations, several velocities profiles along \(x\) were obtained from the results and the viscous stress at these locations were calculated. Such values can then be applied in the set of definitions given from Eqn.~\ref{shear_wall} to Eqn.~\ref{friction_reynolds} following presented in order to calculate \(Re_{\tau}\).

\begin{equation}
{\tau}_w = \rho\nu\left(\frac{d\left<U\right>}{dy}\right)_{y=0}
\label{shear_wall}
\end{equation}

Where \(\rho\) is the density and \({\tau}_w\)  is the shear stress at the wall.

\begin{equation}
u_{\tau} = \sqrt{\frac{{\tau}_w}{\rho}}
\label{u_t}
\end{equation}

Where \(u_{\tau}\) is the friction velocity. And finally,

\begin{equation}
Re_{\tau} = \frac{u_{\tau}\delta}{\nu}
\label{friction_reynolds}
\end{equation}

Where \(\delta\) is the height of the simulated channels. For all Cases studied here this is a constant parameter \(\delta=2\).

The friction Reynolds number calculated using Eqn.~\ref{friction_reynolds} with the numeric simulations' results is then compared with Eqn.~\ref{reynolds_approximation} from Ref.~\cite{pope}, which is valid for fully developed turbulent flows.

\begin{equation}
Re_{\tau} = 0.09Re^{0.88}
\label{reynolds_approximation}
\end{equation}

The Reynolds numbers used to supply Eqn.~\ref{reynolds_approximation} are those varying through \(x\) from Cases I, II and III, as presented in Fig.~\ref{fig:reynolds}. The convolution to be performed over Eqn.~\ref{reynolds_approximation} is given by Eqn.~\ref{convolution}. In this equation \(Re(\chi)\) stands for the analytical approximation given by Eqn.~\ref{reynolds_approximation}, \(g(x-\chi)\)  is a shifting function and \(\chi\) is simply a dummy variable for the streamwise distance.

\begin{equation}
F(x) =  \int_{-\infty}^{+\infty} Re_{\tau}(\chi)g(x-\chi)d\chi
\label{convolution}
\end{equation}

As mentioned earlier, the simulated channels in the present study are divided in three Regions, as shown in Fig.~\ref{fig:geometry_turbchannel}, thus, a different delay should be employed for each region. Since there is no delay in Region I, no special treatment is required for it, thus \(F(x)=Re_{\tau}(x)=550\). In Region II and III the results from the numerical experiments are delayed when compared to those from Eqn.~\ref{reynolds_approximation} and as a result we apply a convolution operator. In both Regions, a decaying exponential has been used as a shifting function, as proposed in Ref.~\cite{signals}. In Region II, the friction Reynolds number is approximated for simplicity as a linear function, given by Eqn.~\ref{region2_reynolds}.

\begin{equation}
Re_{\tau}(x)=ax+550
\label{region2_reynolds}
\end{equation}

where \(a\) is the linear coefficient of the increasing \(Re_{\tau}\) for each one of the three cases in Region II. Using Eqn.~\ref{region2_reynolds} in conjunction to the definition from Eqn.~\ref{convolution}, one may derive Eqn.~\ref{region2_delayed}, which is the delayed function valid for Region II of the considered Cases.

\begin{equation}
F(x)= \frac{a}{R^2}(e^{-R(x-{\delta}_{II})}-1)+\frac{a}{R}(x-{\delta}_{II})+550
\label{region2_delayed}
\end{equation}

Where \({\delta}_{II}\) is the streamwise position that Region II starts and \(R\) is named relaxation parameter and it stands for the exponential decaying constant to be used in the shifting function \(g(x-\chi)\) when dealing with Region II.

Eqn.~\ref{region3_delayed} is the delayed function yielded applying Eqn.~\ref{convolution} over Region III, where a constant friction Reynolds number \(Re_{\tau}\) predicted by Eqn.~\ref{reynolds_approximation} takes place.

\begin{equation}
F(x)= (20000-{\Lambda})(1-e^{-C(x-{\delta}_{III})})+{\Lambda}
\label{region3_delayed}
\end{equation}

In this equation \({\delta}_{III}\) is the streamwise position that Region III starts, \(C\) is named contraction parameter and it stands for the exponential decaying constant to be used in the shifting function \(g(x-\chi)\) when dealing with Region III and finally \({\Lambda}=\frac{a}{R^2}(e^{-R({\delta}_{III}-{\delta}_{II})}-1)+\frac{a}{R}({\delta}_{III}-{\delta}_{II})+550\), represents the contribution from Regions II to the delayed signal in Region III.

In summary, Eqn.~\ref{delayed} express the delayed function for \(Re_{\tau}\) over different Regions of the turbulence channels considered in the present work.

\begin{equation} \label{delayed}
F(x) = \\
    \left\{\begin{array}{@{}lr@{}}
        550  & \text{(Region I)}\\
        \frac{a}{R^2}(e^{-R(x-{\delta}_{II})}-1)+\frac{a}{R}(x-{\delta}_{II})+550 & \text{(Region II)}\\
        (20000-{\Lambda})(1-e^{-C(x-{\delta}_{III})})+{\Lambda} &  \text{(Region III)}
\end{array}\right.
\end{equation}


\section*{RESULTS}

Results for the three cases considered in the present work are shown in this section. First, a comparison between the results in Region I from Case I with existing data from Ref~\cite{myoungkyu2015} is done in order to verify the model. Then, first and second order statistics results in the close to the wall region are presented also just for Case I in the interest of brevity. Third, spatially varying \(Re_{\tau}\) results and a comparison of the results with available correlations are presented. Lastly, visualization of low velocity streaks are provided for \(y^+<10\) region from Case I with a discussion on the evolution of these coherent structures.

\subsection*{Flow visualization}

\begin{figure*}[t]
    \includegraphics[width=\textwidth]{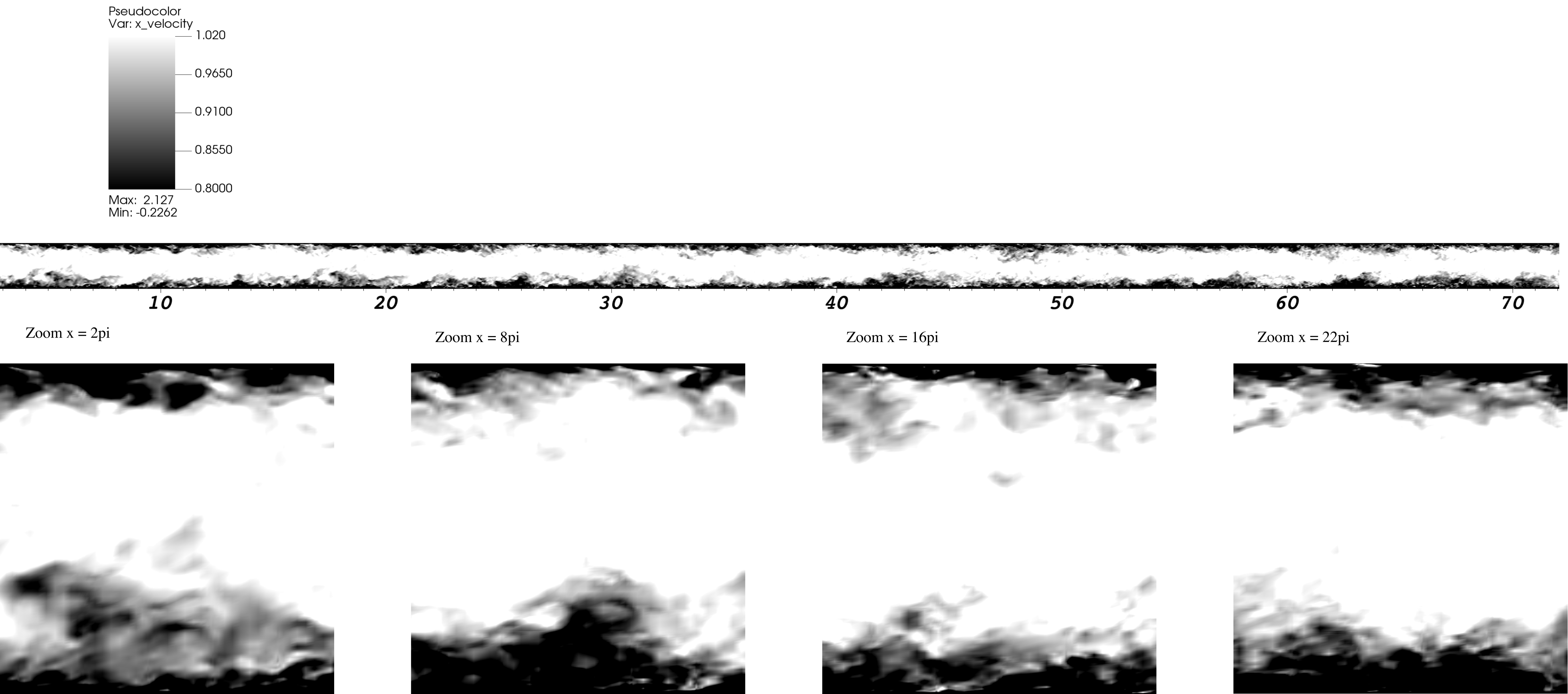}
    \caption{Flow visualization of the streamwive velocity in the y-x plane.}
    \label{fig:flow_viz}
\end{figure*}

Figure~\ref{fig:flow_viz} provides a flow visualization of the streamwise velocity as the flow evolves downstream. The units are normalized by the half-width.

\subsection*{Model verification}

\begin{figure}[t]
\centering
\scalebox{0.5}
{\includegraphics{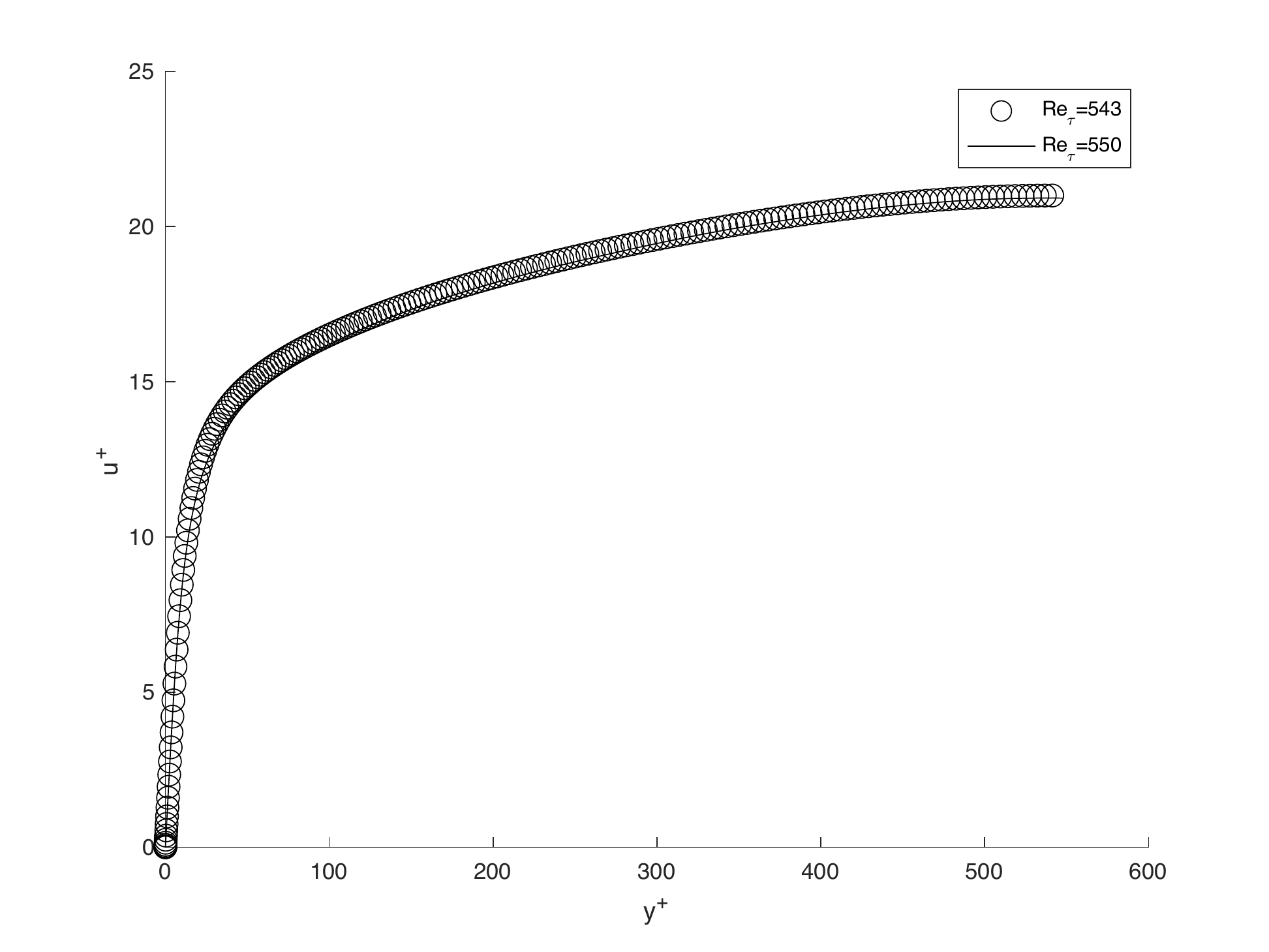}}
\caption{Comparing \(u^+\) from Region I in Case I, where \(Re_{\tau}=550\), with already existing data from Ref.~\cite{myoungkyu2015}.}
\label{fig:comparing_u_543}
\end{figure}

\begin{figure}[t]
\scalebox{0.5}
{\includegraphics{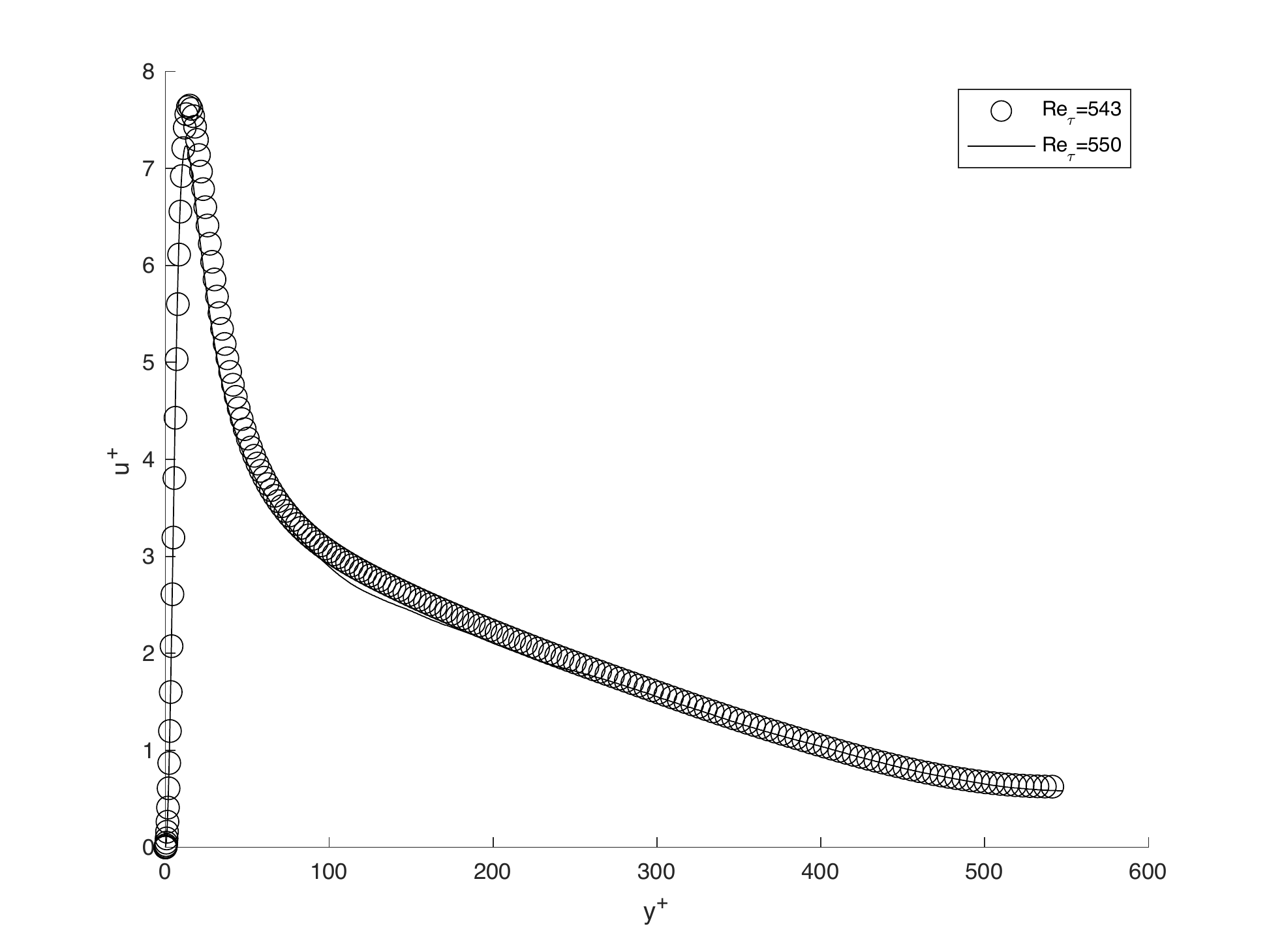}}
\caption{Comparing \(<uu>\) from Region I in Case I, where \(Re_{\tau}=550\), with already existing data from Ref.~\cite{myoungkyu2015}.}
\label{fig:comparing_uu_543}
\end{figure}

In order to verify the model, results from first region in Case I, where \(Re_{\tau}=550\), were compared to existing data from Ref.~\cite{myoungkyu2015}, in which DNS results of a turbulent channel with \(Re_{\tau}=543\) can be found. Fig.~\ref{fig:comparing_u_543} shows the mean velocity \(u^+\) versus \(y^+\) plot, while Fig.~\ref{fig:comparing_uu_543} shows the Reynolds stress \(<uu>\) versus \(y^+\) plot. These graphs show us good agreement between the results and grants us confidence into the models used in the present work. Future work will include comparisons of high order statistics including skewness, kurtosis and flatness as well as turbulence budgets.

\subsection*{First and second order statistics}

The averaging time employed to obtain statistics results for Case I was judged sufficient since it allowed the solution to reach near perfect symmetry. Furthermore, to reduce this time spatial average has been taken over the spanwise direction, since the flow is homogeneous in this direction.

\begin{figure}[t]
\centering
\scalebox{0.5}
{\includegraphics{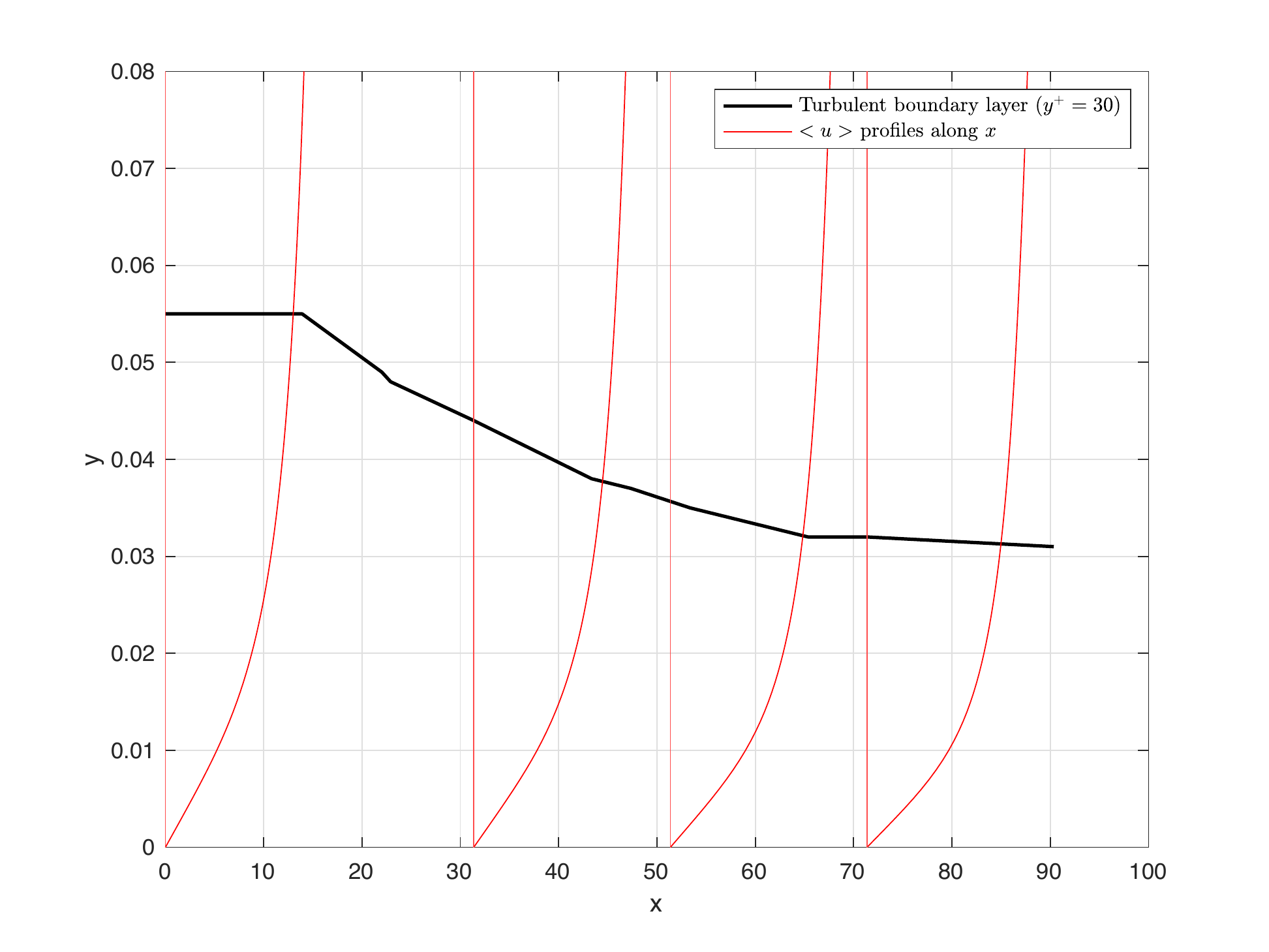}}
\caption{Mean velocity \(<u>\) profiles along the streamwise direction and the turbulent boundary layer \((y^+=30)\).}
\label{fig:u_CI}
\end{figure}

Fig.~\ref{fig:u_CI}  shows how \(<u>\) profile develops along the channel in Case I. From this result we can clearly see that the viscous stress at the wall \(\left(\frac{d\left<U\right>}{dy}\right)_{y=0}\) gets more pronounced as we move in the streamwise direction. This result is consistent with expectations since in Region II the viscosity starts decreasing linearly, as shown through Fig.~\ref{fig:viscosity}, causing the viscous effects to become more pronounced. This fact also explains why the turbulent boundary layer \((y^+=30)\) becomes closer to the wall in this same range, fact also presented in Fig.~\ref{fig:u_CI}. Moreover, after a certain value of \(x\), the boundary layer reaches a constant height from the wall, this fact also makes sense once in Region III the viscosity becomes constant, not causing a change in the viscous stress anymore.

\begin{figure}[t]
\centering
\scalebox{0.5}
{\includegraphics{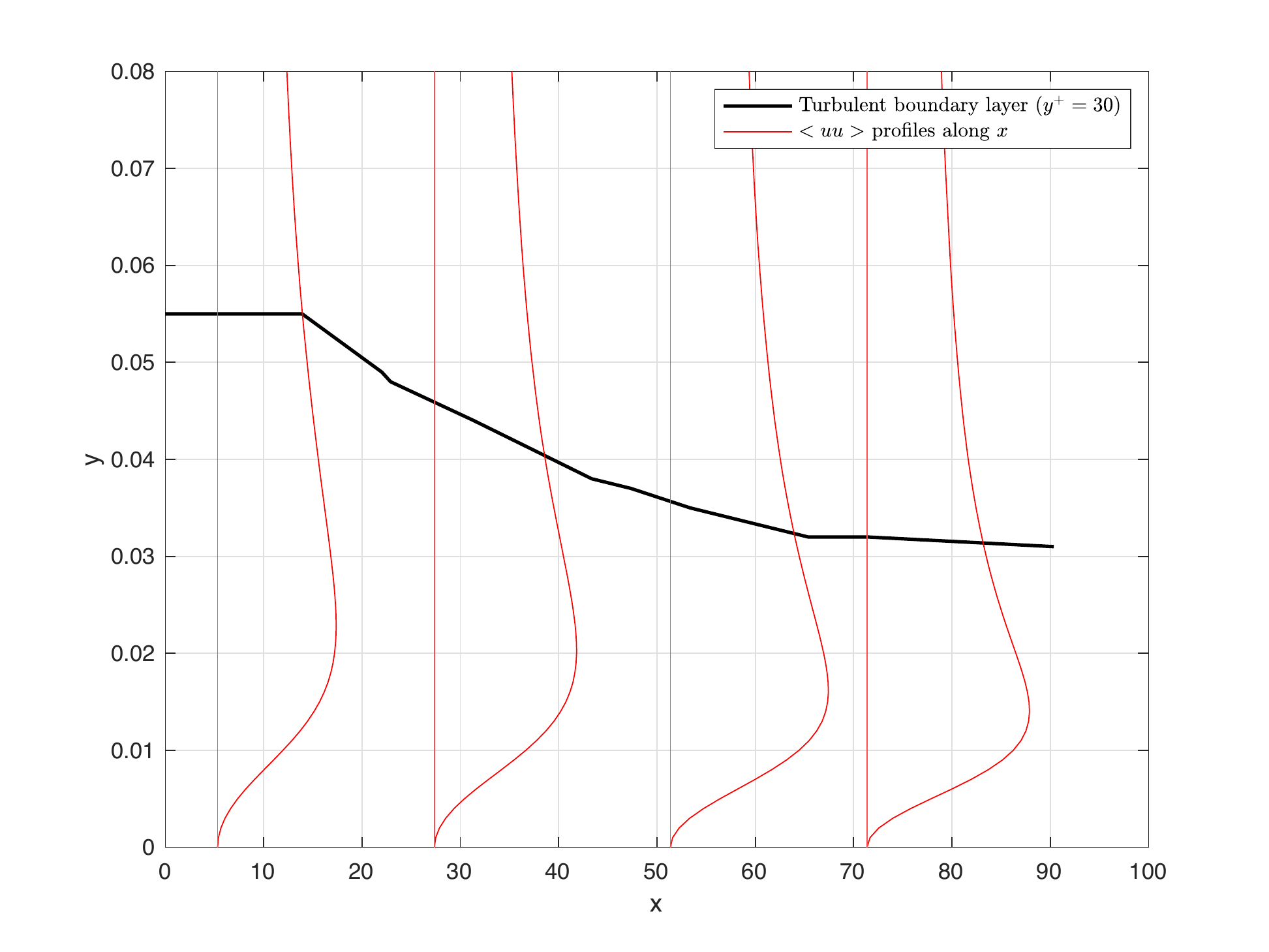}}
\caption{Reynolds stress \(<uu>\) profiles along the streamwise direction and the turbulent boundary layer (\(y^+=30)\).}
\label{fig:uu_CI}
\end{figure}

Fig.~\ref{fig:uu_CI} shows the development of the profile of \(<uu>\) in the streamwise direction also for Case I. As we move forward in the streamwise direction in this plot, the peak of \(<uu>\) increases and gets closer to the wall. This result is also consistent with expectations.

\subsection*{Delayed friction Reynolds number}

In practice terms, Eqn.~\ref{delayed} delays the signal resulted from the approximation given by Eqn.~\ref{reynolds_approximation}. This is true once the former expression is the result of a convolution operation of the last one. One might notice that Eqn.~\ref{delayed} depends on two parameters, i.e., the contraction parameter \(C\) and the relaxation parameter \(R\). We adjust these parameters until the delayed function \(F(x)\) fits well with the simulation results.

\begin{figure}[t]
\centering
\scalebox{0.42}
{\includegraphics{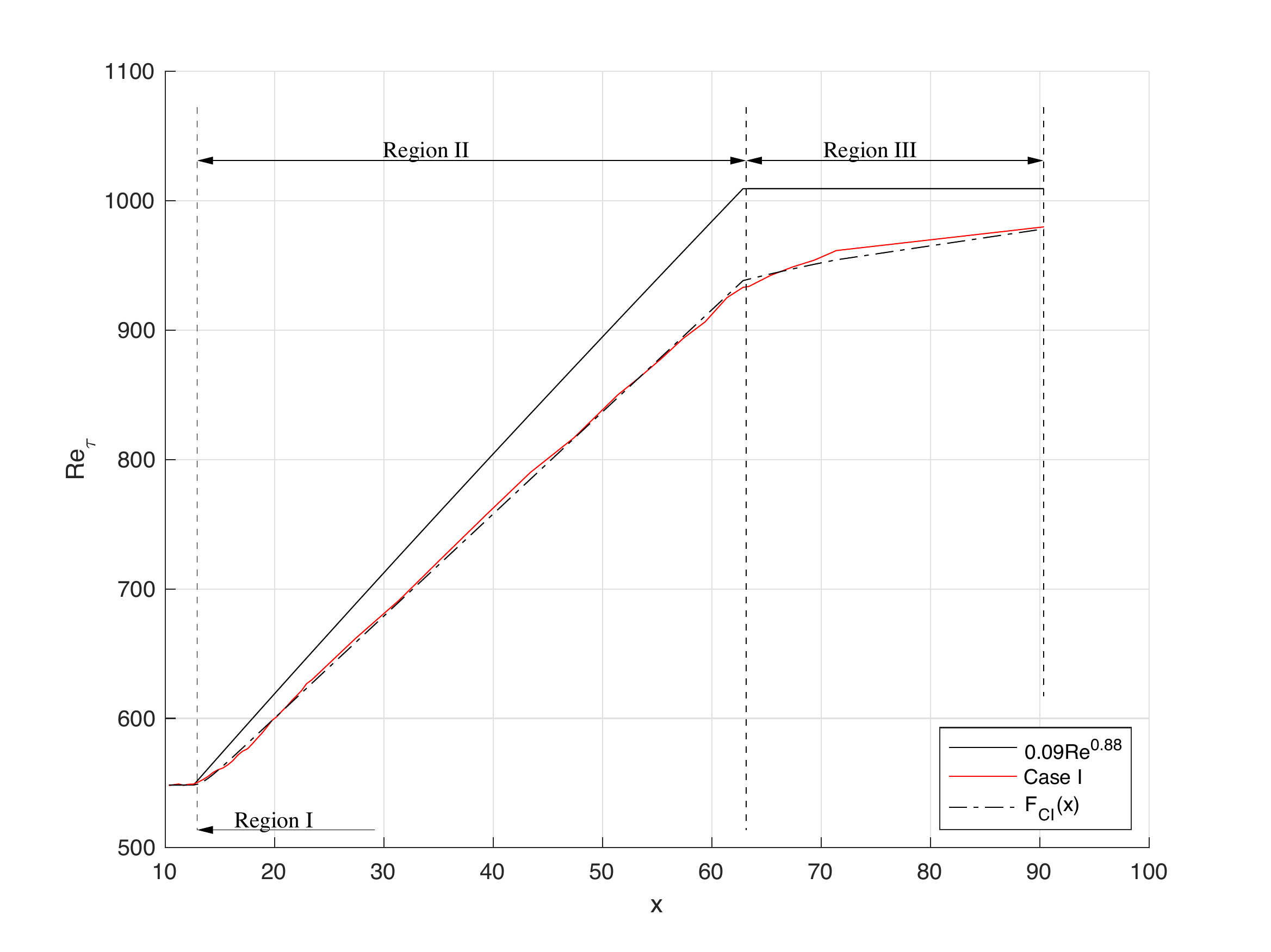}}
\caption{Friction Reynolds number through \(x\) for Case I.}
\label{fig:convolution_CI}
\end{figure}

\begin{figure}[t]
\centering
\scalebox{0.5}
{\includegraphics{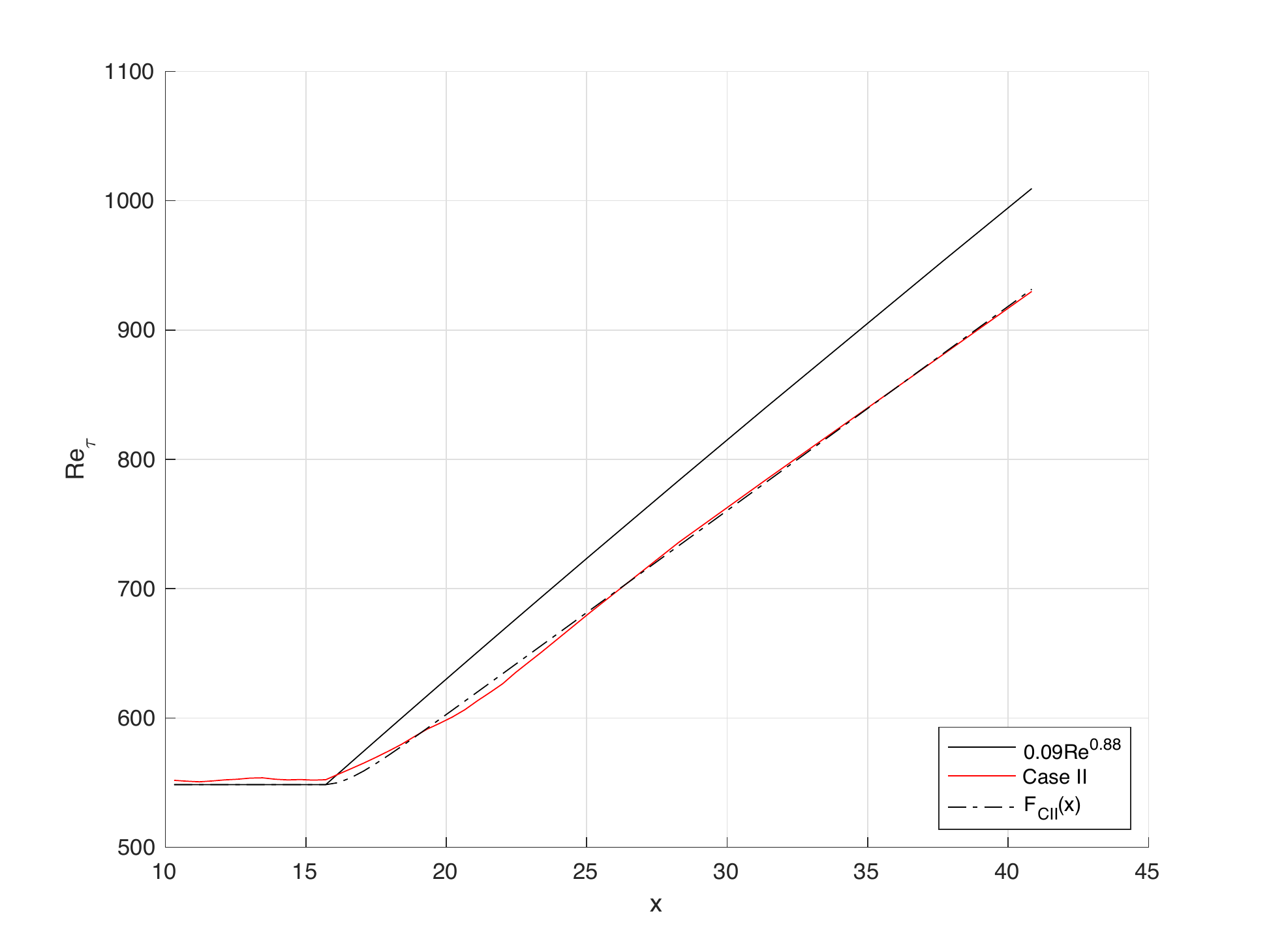}}
\caption{Friction Reynolds number through \(x\) for Case II.}
\label{fig:convolution_CII}
\end{figure}

\begin{figure}[t]
\centering
\scalebox{0.5}
{\includegraphics{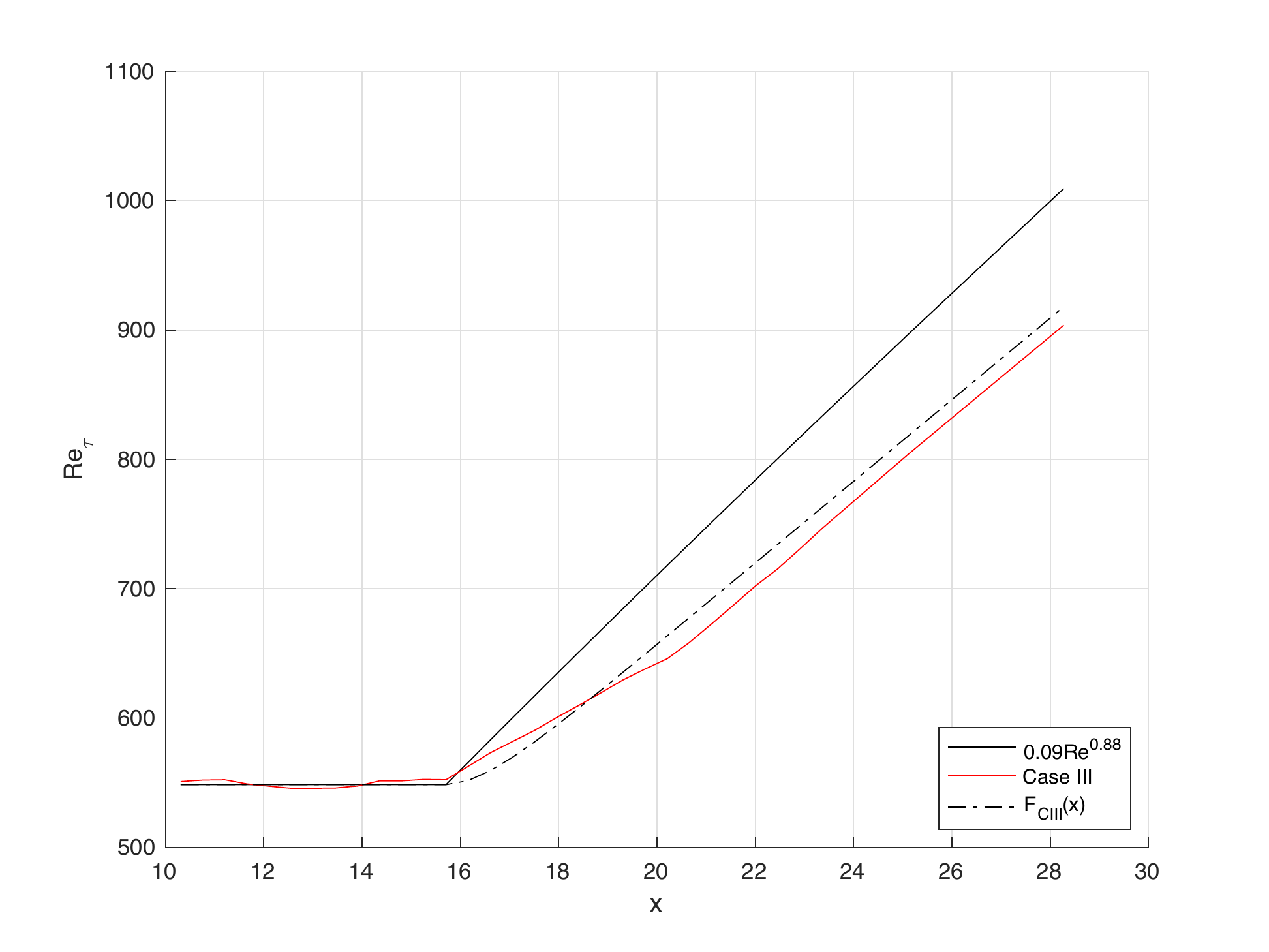}}
\caption{Friction Reynolds number through \(x\) for Case III.}
\label{fig:convolution_CIII}
\end{figure}

The above procedure has been performed for Cases I, II and III considered in the present work. Figs.~\ref{fig:convolution_CI}-\ref{fig:convolution_CIII} provide the results for the three simulations. In each one of these plots the parameters \(C\) and \(R\) were calibrated so the delayed function fitted to the simulation results. The estimated values for these two parameters in Case I are \(C=1.162\) and \(R=0.030\). One should notice that \(C\) does not change between the three cases regardless how fast the viscosity changes along \(x\). We note that the \(R\) value may however depend on the ratio between maximum Reynolds number and minimum Reynolds number for the ramp, which has been kept constant. The potential effect of the ramp on \(R\) will be further investigated in future studies.

\begin{figure}[t]
\centering
\scalebox{0.5}
{\includegraphics{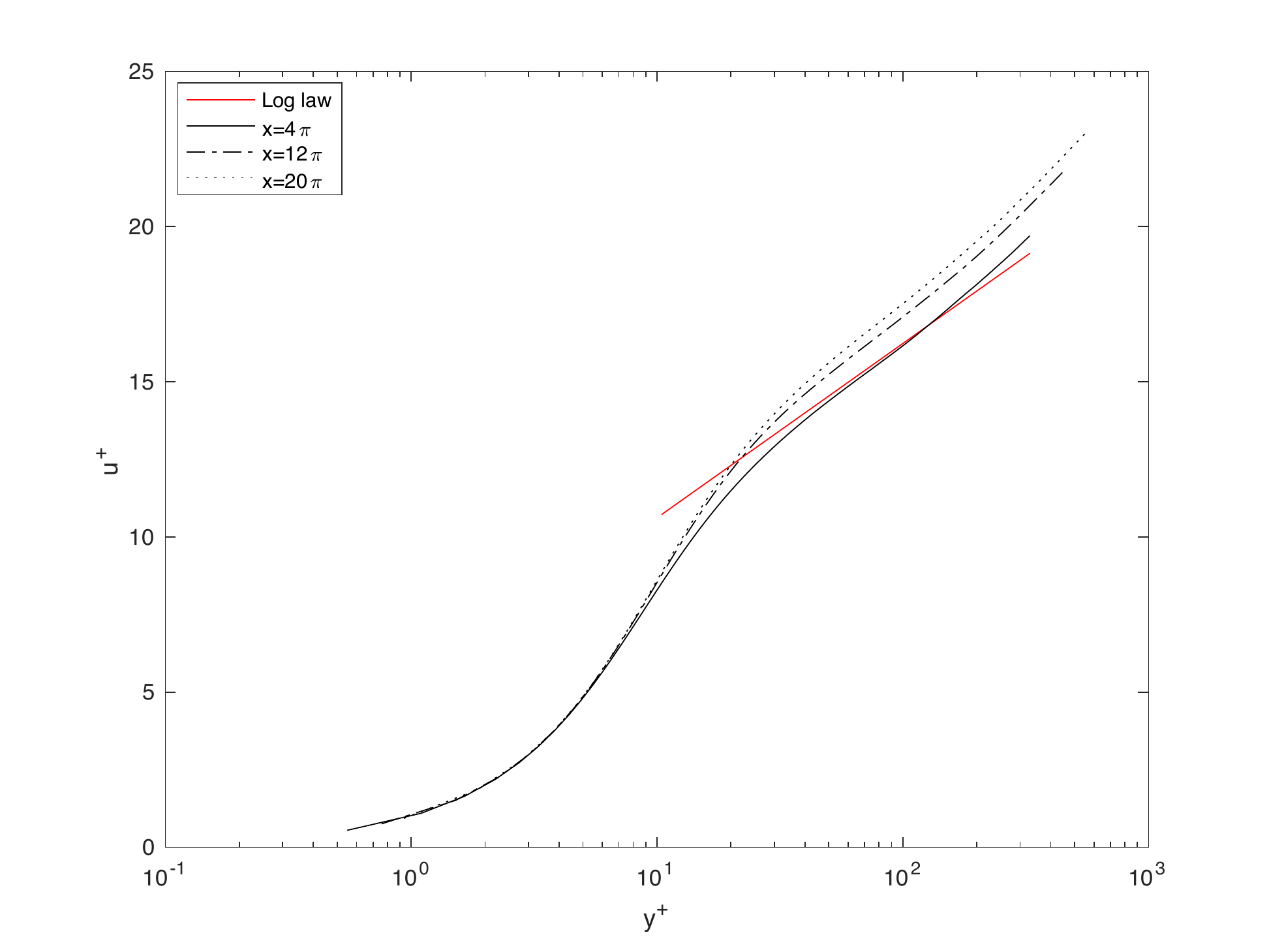}}
\caption{\(u^+\) versus \(y^+\) obtained from Case I at \(x=4{\pi}\) (end of Region I), \(x=12{\pi}\) (middle of Region II) and \(x=20{\pi}\) (end of Region II) compared to the log law.}
\label{fig:log_law_CI_upstream}
\end{figure}

\begin{figure}[t]
\centering
\scalebox{0.5}
{\includegraphics{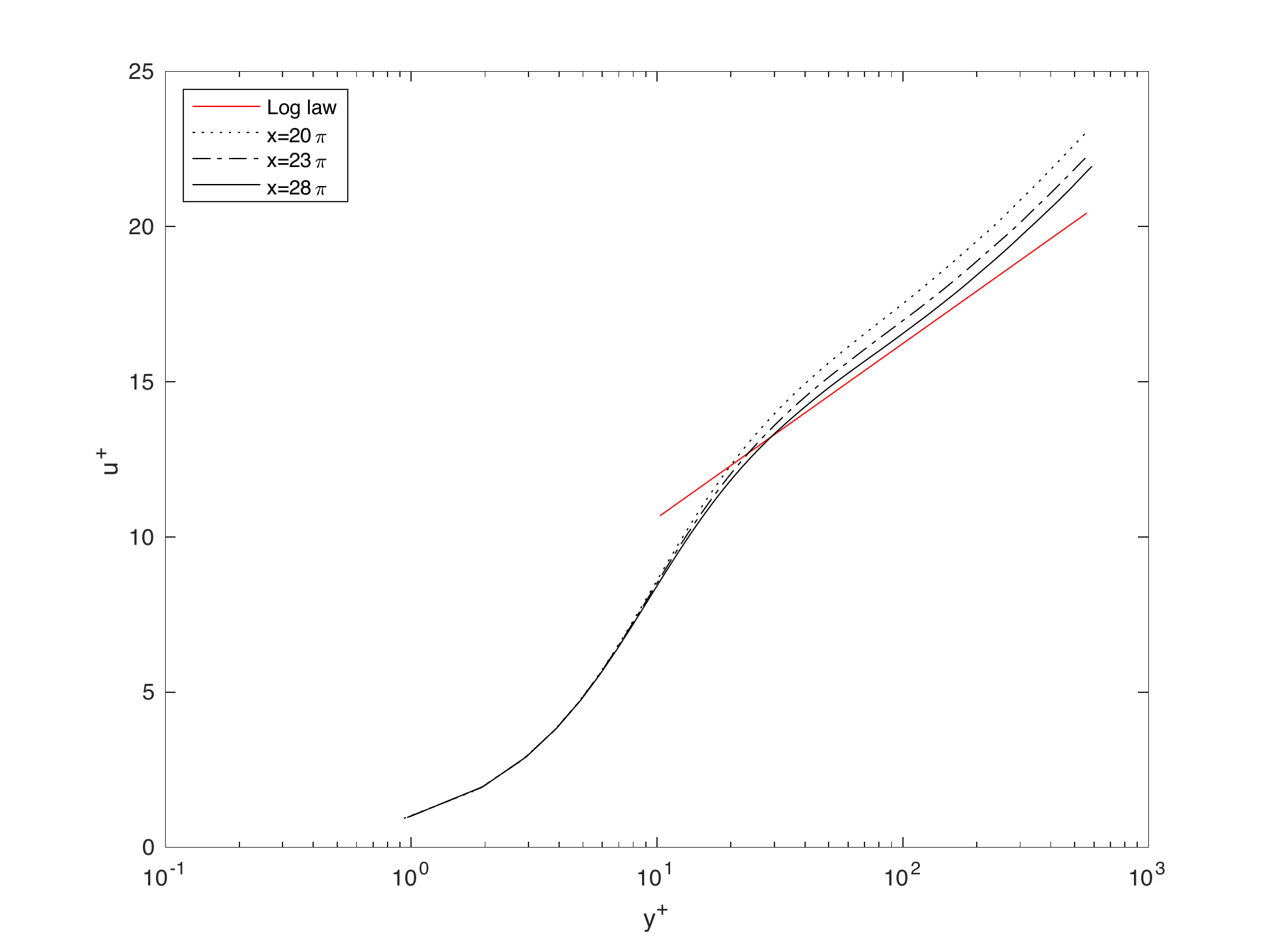}}
\caption{\(u^+\) versus \(y^+\) obtained from Case I at \(x=20{\pi}\) (end of Region I), \(x=23{\pi}\) (Region III) and \(x=28{\pi}\) (Region III) compared to the log law.}
\label{fig:log_law_CI_downstream}
\end{figure}

Fig.~\ref{fig:log_law_CI_upstream} and Fig.~\ref{fig:log_law_CI_downstream} shows a comparison between the mean velocity profile \(u^+\) versus \(y^+\) from Case I and the log-law at different locations along \(x\).

In Fig.~\ref{fig:log_law_CI_upstream} we can see that the \(u^+\) profile is in good agreement with the log-law at \(x=4{\pi}\), however, insofar we move in the downstream direction, the profile departures from the law, which are the cases at \(x=12{\pi}\) and further at \(x=20{\pi}\). This departure is consistent with observations made in the graph Fig.~\ref{fig:convolution_CI}. We may notice that the departure from the log-law follows the departure of the friction Reynolds number obtained from Case I to the one using Eqn~\ref{reynolds_approximation}. This fact becomes more clear when analyzing the profile at \(x=20{\pi}\), this is the location where \(Re_{\tau}\) delays the most from a fully developed flow and, for this reason, departures the most from the log-law.

In Fig.~\ref{fig:log_law_CI_downstream} the opposite trend from Fig.~\ref{fig:log_law_CI_upstream} is observed, i.e., as we move in the downstream direction throughout Region III, \(u^+\) profile tends to become closer to the log-law again. This fact is closely related to what is observed in Fig.~\ref{fig:convolution_CI} as well. From Fig.~\ref{fig:convolution_CI} we observe that \(Re_{\tau}\) in Region III approaches the value expected for fully developed turbulent flows.

\subsection*{Turbulent structures}

\begin{figure*}[t]
	\includegraphics[width=\textwidth]{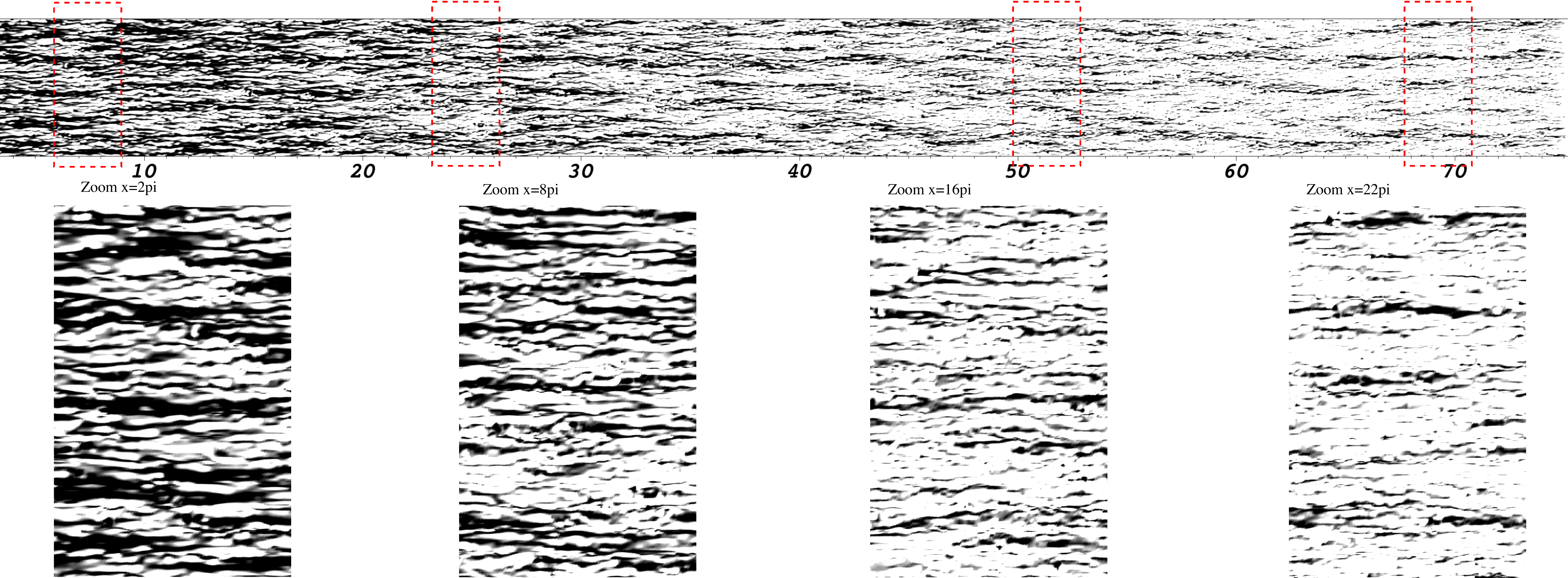}
	\caption{A cross section of the instantaneous streamwise velocity field at y=0.01  from the wall.}
	\label{fig:streaks}
\end{figure*}

\begin{figure}[t]
\centering
\scalebox{0.5}
{\includegraphics{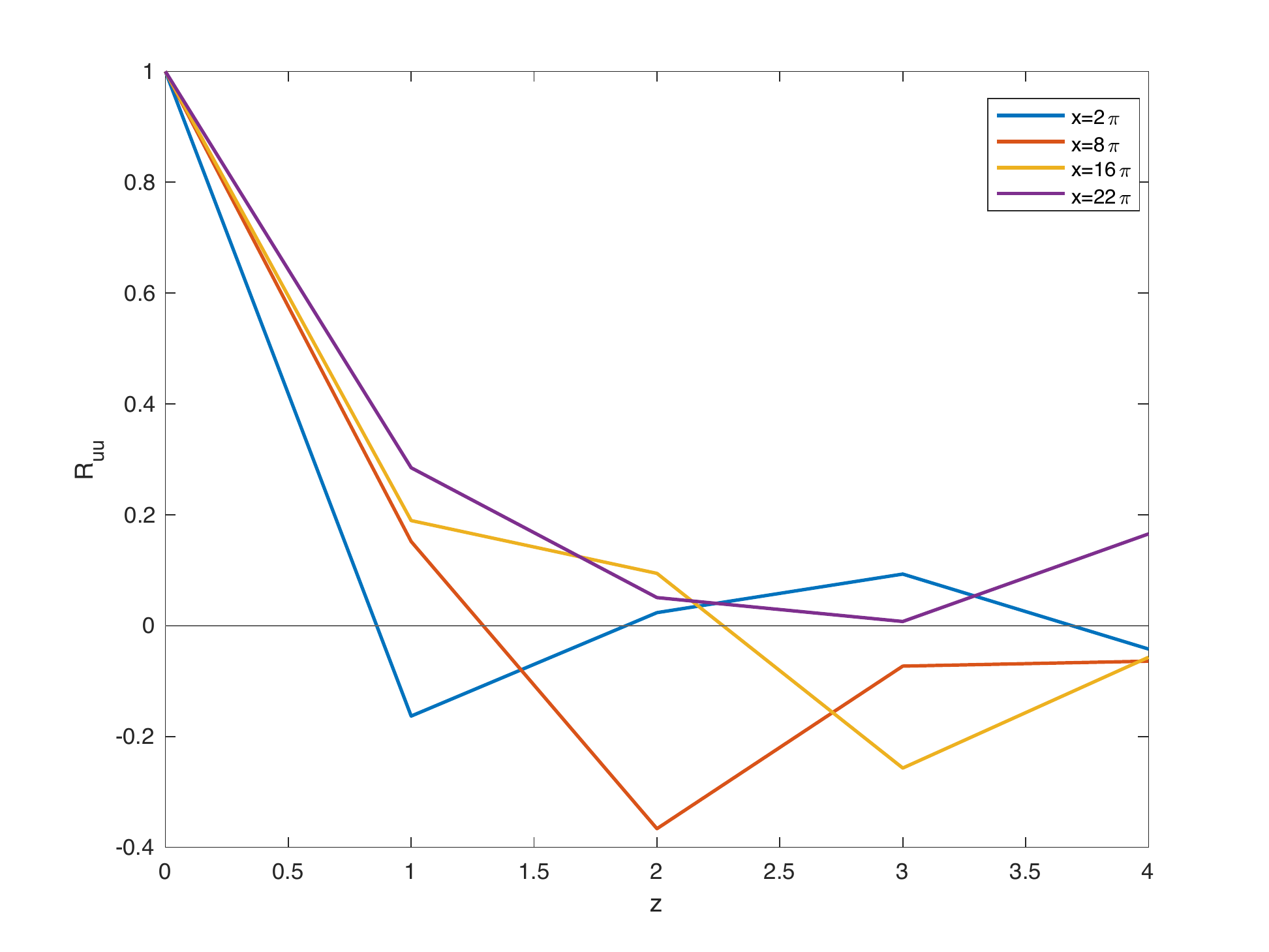}}
\caption{Two-point correlation of \(u\) over the spanwise direction at different streamwise distances on a plane \(y=0.01\) from the wall.}
\label{fig:spanwise_correlation}
\end{figure}

\begin{figure}[t]
\centering
\scalebox{0.5}
{\includegraphics{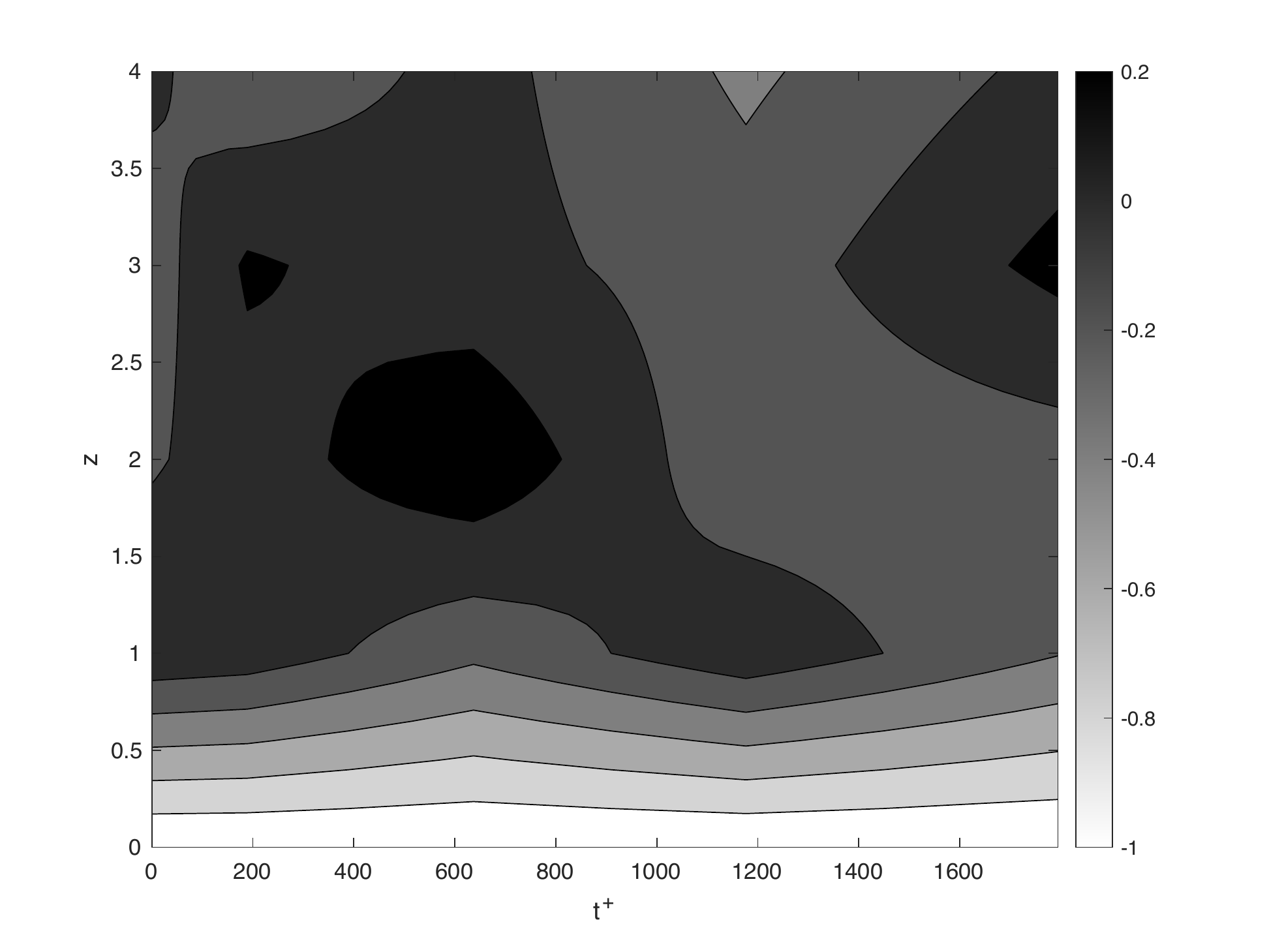}}
\caption{Time variation of the spanwise correlation of the streamwise velocity on a plane \(y=0.01\) from the wall. The dimensionless time scale is \(t^+=tu_{\tau}^2/{\nu}\).}
\label{fig:contour_map}
\end{figure}

Fig.~\ref{fig:streaks} depicts the  instantaneous streamwise velocity scalar field at a height \(y=0.01\) from the wall. Low-velocities streaks are observed. Through visual verification, it is possible to measure that the length of these structures varies in a range of \(500\) to \(2000\) wall units, which is consistent with the literature Ref.~\cite{carlier2005}. Furthermore, it is also noticed that the streaks' length does not change along the streamwise direction. This was confirmed through streamwise two-point correlations at multiple locations. In the future we will also apply advanced streak length measurement techniques.

Two-points correlation in the spanwise direction has been performed correlating the streamwise velocity for the present work. These measures has been taken over different positions in the turbulence channel in Case I, Fig.~\ref{fig:spanwise_correlation} shows the results. One may notice that as we move forward in \(x\) the correlation of the measured signals becomes more strong until it reaches a peak in negative value. The peak negative value can be used as a mean to evaluate correlation and separation of the streaks.  The increasing correlation is consistent with what seen  in Figure~\ref{fig:streaks}, i.e., these structures becomes more separated as we move in the downstream portion of the channel.

Results from Fig.~\ref{fig:spanwise_correlation} were mapped into a dimensionless time scale using the Taylor hypothesis generating Fig.~\ref{fig:contour_map}, where the time variation of the spanwise correlation of the streamwise velocity is presented. This allows for a direct comparison with transient channel flow. In fact the results indicates that streaks are destroyed and destabilized as the flow advances  in the streamwise direction. The scenario is remarkably similar to what presented by He \& Seddighi ~\cite{he2015} in transient channel flow at high to moderate Reynolds ratios.  We note that the behavior may change with different Reynolds ratios as it will be investigated in the future.

\subsection*{RANS simulation}

\begin{figure}[t]
\centering
\scalebox{0.5}
{\includegraphics{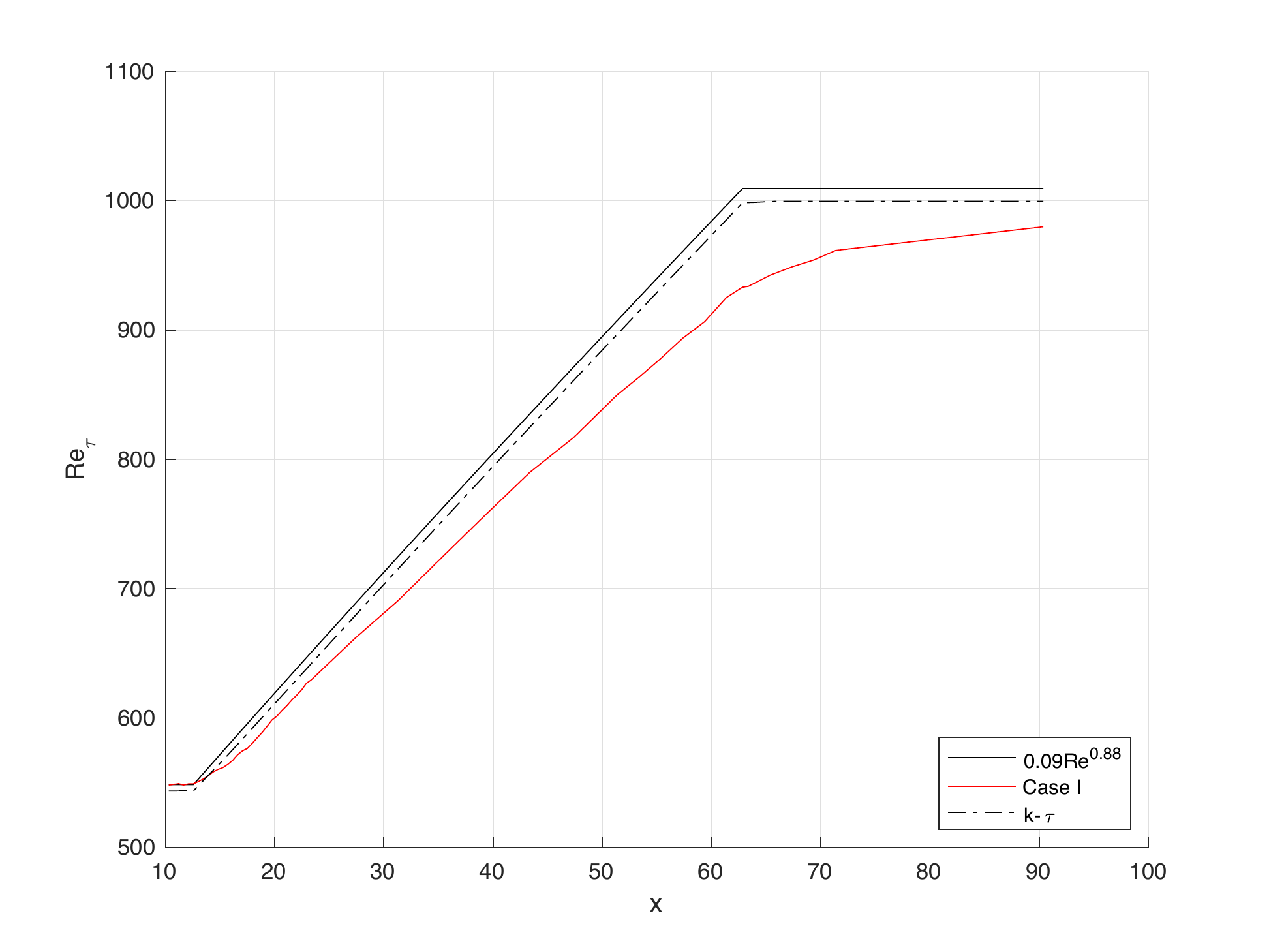}}
\caption{Friction Reynolds number through \(x\) for Case I using \(k-\tau\) model.}
\label{fig:Re_RANS}
\end{figure}

In order to check the applicability of standard turbulence models for varying viscosity flows, an analysis using \(k-\tau\) model was performed for Case I. An 8th polynomial order simulation was performed. The RANS implementation follows what outlined in Ref.~\cite{Tomboulidesa2018}. Fig.~\ref{fig:Re_RANS} presents the fiction Reynolds number resulted from this simulation.

The RANS simulation gives us a \(Re_{\tau}\) of 544 and 1000 for Region I and Region III respectively, i.e., values that are close compared to Eqn.~\ref{reynolds_approximation}, with a difference less than 0.009\%. In Region II Eqn.~\ref{reynolds_approximation} increases quasi linearly, featuring the same behaviour of Eqn.~\ref{reynolds_approximation}, thus, not being able to predict the delay found in the DNS results. This result calls into question the application of standard turbulence models, e.g., the \(k-\tau\) model, in cases where viscosity changes significantly over the domain.

\section*{CONCLUSIONS}

In the present work, DNS simulation has been performed in order to analyze the effect of varying viscosity in turbulence channel flow. Three cases were considered, in each one of them the viscosity varied along the channel with different ramps, causing the Reynolds number to range linearly from \(10,000\) to \(20,000\) in the ramp region. The benchmark is relevant to nuclear applications with step changes in viscosity.

The friction Reynolds number was computed for each one of the three considered cases using DNS and for Case I using \(k-\tau\) model. By comparing these results with an existing approximation valid for fully developed turbulent flows from Ref.~\cite{pope} we noticed a delay in the development of the wall shear. This was consistent with expectations. While, the viscosity affects the viscous stress at the wall boundaries, turbulence dynamics does not respond instantaneously. This same feature was also observed by several authors for transient channel flow ~\cite{maruyama1976,he2000,greenblatt2004,he2015}. Another interesting finding is that the measured delay in \(Re_{\tau}\) does not depend on the inclination of the Reynolds number ramp. This behavior is also consistent with transient channel flow.

Furthermore, from both visual inspection and two-point correlation it was noticed that the distance between the streaks in the spanwise direction appeared in the near-wall region to increase until it reaches a peak. This is consistent with trends observed by other authors for transient channel flow for an appropriate range of Reynolds ratios. We note that while we explored different ramps we did not investigate different Reynolds ratios.

Overall, we observed strong similarities between the flow presented here and transient channel flow. This opens the potential to use the rich theory developed for the class of these flows to this geometry. This possibility will be explored in the future.  Further investigation will also be dedicated to better describe and understand the behavior of the low-speed streaks in the near wall region. We propose using image processing techniques to analyze structures in this region. To achieve this goal we may refer to image procedures allowing quantitative characterization of structures in turbulent flows, as proposed in Ref.~\cite{lin2008}, allied to machine learning techniques. We will also  perform coupled temperature-velocity calculations with variable properties to assess the effect of temperature transport over the behavior of turbulence.

Finally, we will compute budget terms for the Reynolds stress tensor with the objective of providing a stronger benchmark basis for RANS models.  In fact, results presented here call into question the application of standard turbulence models in cases where viscosity changes significantly over the domain (i.e., a two-fold reduction). The availability of budget terms for the Reynolds stress tensor will be the starting point to assess in more depth multiple RANS models and eventually suggest modifications if needed.

\bibliographystyle{asmems4}

\bibliography{asme2e}
\appendix

\end{document}